\documentclass[twocolumn]{aastex631}
\usepackage{amsmath}
\usepackage{mathrsfs}
\usepackage{CJK}
\usepackage{multirow}

\shorttitle{The IMBH}
\shortauthors{Xia et al.}

\graphicspath{{./}{figures/}}

\begin{document}
\begin{CJK*}{UTF8}{gbsn}

\title{How do the LIGO-Virgo-KAGRA's Heavy Black Holes Form? No evidence for core-collapse Intermediate-mass black holes in GWTC-4}

\author[0000-0002-2630-4753]{Fan-Xiao-Yu Xia (夏凡小雨)}\affiliation{Key Laboratory of Dark Matter and Space Astronomy, Purple Mountain Observatory, Chinese Academy of Sciences, Nanjing 210023, People's Republic of China}

\author[0000-0001-9626-9319]{Yuan-Zhu Wang (王远瞩)}\affiliation{Institute for Theoretical Physics and Cosmology, Zhejiang University of Technology, Hangzhou, 310032, People's Republic of China}\email{The corresponding author: vamdrew@zjut.edu.cn (Y.Z.W)}

\author[0000-0002-2956-8367]{Ying Qin (秦颖)}\affiliation{Department of Physics, Anhui Normal University, Wuhu, Anhui 241000, People’s Republic of China}

\begin{abstract}
We investigate the population properties of binary black holes (BBHs) from the LIGO–Virgo–KAGRA collaboration, focusing especially on those in the high-mass range, using the newly released GWTC-4 catalog. For the first time, we search for a subpopulation of low-spin intermediate-mass black holes (IMBHs) that would indicate formation via stellar core collapse. With the currently available catalog, we find no evidence for such a subpopulation, and set a 90\% upper limit on the merger rate of collapse-formed IMBHs at $0.077~\mathrm{Gpc}^{-3}\,\mathrm{yr}^{-1}$. The mass distribution of low-spin (stellar-origin) black holes truncates at $65^{+23}_{-22}\,M_\odot$, consistent with the lower edge of the pair-instability mass gap (PIMG), although we cannot directly determine its upper boundary from current data. Informed by stellar evolution theory, we estimate the upper edge of the PIMG to be $150\pm24\,M_\odot$. We find that the observed IMBHs belong to a high-spin subpopulation, consistent with formation through successive hierarchical mergers. 

\end{abstract}


\section{Introduction}

LIGO, Virgo, and KAGRA \citep{2015CQGra..32g4001L,2015CQGra..32b4001A,2013PhRvD..88d3007A} have detected more than 300 gravitational-wave (GW) signals from binary black hole (BBH) mergers in the first four observing runs (see \href{https://gracedb.ligo.org/superevents/public/O4/}{GraceDB}), which have contributed to foudmantal Physics and Cosmology \citep{2017Natur.551...85A,2025arXiv250902047W,2025PhRvL.135k1403A,2025ApJ...993L..21A,2026SciBu..71...83T}. A significant fraction of BBHs have unexpectedly large masses, challenging stellar evolution theories \citep{2021ApJ...913L...7A,2023PhRvX..13a1048A,2025arXiv250818083T}.

Intermediate-mass black holes \citep[IMBHs,][]{2004IJMPD..13....1M}, with masses ranging from $\sim 10^{2}$ to $10^{5}\,M_\odot$, have previously been observed as remnants of BBH mergers \citep[e.g., GW190521][]{2020PhRvL.125j1102A}. Recent observations show that some IMBHs are also merging with other BHs \citep[e.g., GW231123][]{2025arXiv250708219T}. 

These unexpected events have sparked discussion regarding the origin and evolution of IMBHs in BBH systems \citep[e.g.,][]{2025ApJ...992L..26S,2025arXiv250717551L,2025arXiv250715701Y,2025arXiv250809965D,2025arXiv250810088C,2025arXiv250808558B,2025ApJ...995L..76P,2025ApJ...994L..37K,2025arXiv251217631G,2025ApJ...994L..54P}.

Stellar evolution models predict an upper mass cutoff of $\sim 40$--$65\,M_\odot$ for BHs formed via core collapse, owing to (pulsational) pair-instability supernovae \citep[(P)PISN,][]{1964ApJS....9..201F,1967PhRvL..18..379B,2017ApJ...836..244W,2021ApJ...912L..31W}. 
More massive BBHs are therefore expected to form predominantly through hierarchical mergers of stellar-mass BHs in dense environments \citep{2021NatAs...5..749G}. 
Alternatively, the direct collapse of very massive stars---such as Population III stars with helium core masses $\gtrsim 135\,M_\odot$---may also produce IMBHs and contribute to the observed BBH population \citep[e.g.,][]{2016A&A...588A..50M}.

While individual, unusual events may offer clues to BBH formation channels \citep[e.g.,][]{2020PhRvD.102d3015A,2020ApJ...896L..44A,2020PhRvL.125j1102A,2024ApJ...970L..34A,2025arXiv250708219T}, population-level analyses provide a more systematic avenue to identify BBH subpopulations and their corresponding evolutionary pathways \citep{2019ApJ...882L..24A,2021ApJ...913L...7A,2023PhRvX..13d1039A}. 
Different formation channels are expected to imprint characteristic signatures on the BBH population-particularly in component masses and spins-allowing them to be statistically disentangled.

The detection of GW231123 is particularly intriguing, as the system likely contains one component beyond the pair-instability mass gap (PIMG) \citep{2025arXiv250708219T}. 
However, the unusually large spin magnitudes inferred for this event suggest that both components are likely remnants of previous BBH mergers rather than first-generation BHs \citep{2025arXiv250717551L,2025arXiv250715967S,2025arXiv250908298L}. 
Motivated by this event and the full GWTC-4 catalog, we aim to determine whether any evidence exists for IMBH mergers formed via direct stellar core collapse.

Previous population studies indicate that BBHs with component masses above $\sim 45\,M_\odot$ are broadly consistent with hierarchical merger origins \citep{2022ApJ...941L..39W,2024PhRvL.133e1401L,2024ApJ...977...67L,2025ApJ...987...65L,2024ApJ...975...54G,2025PhRvL.134a1401A,2025arXiv250915646B,2025arXiv250904151T,2026arXiv260107908P,2026ApJ...996..144G,2025arXiv251014363P}. While
\citet{2025arXiv251022698W} showed that the maximum-mass cutoff for the low-spin subpopulation may extend up to $\sim 65\,M_\odot$. 
These findings naturally raise several further questions: Are any high-mass BBH events (e.g., those containing IMBH components) formed via stellar collapse? How massive can IMBHs become through hierarchical mergers? And where is the far-side edge of the PISN mass gap, if it exists?

Although recent GWTC-4 population analyses suggest that high-mass events are consistent with hierarchical mergers \citep{2025arXiv250904637A,2025arXiv250915646B,2025arXiv250923897L,2025arXiv251022698W,2025arXiv251105316T}, these studies typically model only two broad subpopulations, making it difficult to identify a distinct low-spin population within or beyond the PIMG. 
Since low-spin events are unlikely to arise from hierarchical mergers \citep{2017ApJ...840L..24F,2017PhRvD..95l4046G,2021NatAs...5..749G}, the detection of such a subpopulation---if present---would provide important insights into primordial BHs or collapse-formed IMBHs. 
In this work, we therefore perform a dedicated population analysis of GWTC-4 aimed at searching for such a population. Additionally, we also aim to characterize the shape of the expected PIMG.

\section{Methods}\label{sec:method}

We perform a population analysis of the BBHs in GWTC-4, employing hierarchical Bayesian inference to estimate the hyperparameters of the population model; see Appendix~\ref{app:meth} for details. 
Following \citet{2025arXiv250818083T}, we select 153 BBH events with a false alarm rate (FAR) $< 1\,{\rm yr}^{-1}$ for our analysis. 
Posterior samples for these events are obtained from \href{https://zenodo.org/doi/10.5281/zenodo.16911563}{events-zenodo} \citep{ligo_scientific_collaboration_2025_16911563}.

\subsection{spin versus mass model (Main Model)}
The spin magnitudes of black holes are among the key observables for determining their origins \citep{2022PhR...955....1M}.
First-generation (core-collapse) BHs and primordial BHs are expected to be either slowly spinning \citep{2019ApJ...881L...1F,2017PTEP.2017h3E01C} or to possess moderate spins acquired through tidal spin-up \citep[e.g.][]{2018A&A...616A..28Q,2016A&A...588A..50M,2020A&A...635A..97B,2026A&A...708A..62W} or accretion \citep[e.g.][]{2022ApJ...930...26S,2022PhR...955....1M} (although see e.g. \citet{2025arXiv250810088C,2025ApJ...994L..37K,2025arXiv250900154P,2025arXiv250808558B} for alternative views).
Higher-generation BHs, in contrast, are expected to be highly spinning, with spin magnitudes that typically peak at $\sim 0.7$ \citep{2017PhRvD..95l4046G,2017ApJ...840L..24F,2021NatAs...5..749G} or even larger values due to gas hardening in AGN disks \citep{2024A&A...685A..51V}.
In this work, we therefore construct a population model designed to identify the subpopulations of higher-generation and first-generation BHs mainly based on their spin magnitudes.

We adopt the population model of \cite{2024PhRvL.133e1401L}, in which the joint distribution of component mass $m$ and spin magnitude $\chi$ is described by a two-component mixture:
\begin{equation}\label{twopop}
\pi(m,\chi|\Lambda)=\pi_1(m,\chi|\Lambda_1)\,(1-r_2) + \pi_2(m,\chi|\Lambda_2)\,r_2,
\end{equation} 
where each subpopulation factorizes into independent mass and spin distributions,
\begin{equation}
\pi_i(m,\chi|\Lambda_i)=P_{m,i}(m|\Lambda_i)\,P_{\chi,i}(\chi|\Lambda_i).
\end{equation}
The spin distribution $P_{\chi,i}(\chi|\Lambda_i)$ is modeled as a Gaussian $\mathcal{G}_{[\chi_{{\rm min},i},\chi_{{\rm max},i}]} (\chi | \mu_{\chi,i},\sigma_{\chi,i})$ truncated to the interval $[\chi_{{\rm min},i},\chi_{{\rm max},i}]$. For the second subpopulation, the mass function is given by a PowerLaw+Spline model $\mathcal{PS}$ \citep{2022ApJ...924..101E}; see Appendix~\ref{app:models} for its definition.
In this work, we extend the mass function of the first subpopulation to allow for a potential contribution from collapse-formed IMBHs beyond the pair-instability mass gap. Specifically, we set
\begin{equation}
\begin{aligned}
&P_{m,1}(m|\Lambda_1)=\mathcal{PS}(m|\Lambda_1)\,(1-r_{\mathcal{IM}}) + \\
&\mathcal{PL}(m|\alpha_{\mathcal{IM}}, m_{{\rm min}, \mathcal{IM}}, m_{{\rm max}, \mathcal{IM}})\,r_{\mathcal{IM}},
\label{equation3}
\end{aligned}
\end{equation}
where $\mathcal{PL}$ is a truncated power-law distribution \citep{2021ApJ...913L...7A}, designed to capture the possible low-spin IMBH population formed via stellar collapse.

The overall population model is 
\begin{equation}
\begin{aligned}
&\pi(\lambda|\Lambda) \propto \pi(m_1,\chi_1,|\Lambda)\pi(m_2,\chi_2,|\Lambda) \mathcal{F}_{\rm pair}(m_1,m_2|\beta)\\
&\times\mathcal{GU}(\cos\theta_1,\cos\theta_2|\zeta,\mu_{\rm t},\sigma_{\rm t})P_z(z|\gamma),
\end{aligned}
\end{equation} 
where $\mathcal{F}_{\rm pair}(m_1,m_2|\beta)=(m_2/m_1)^\beta$ is the pairing function, $\mathcal{GU}(\cos\theta_1,\cos\theta_2|\zeta,\mu_{\rm t},\sigma_{\rm t})=(1-\zeta)U(\cos\theta_1,\cos\theta_2|-1,1)+\zeta \mathcal{G}_{[-1,1]} (\cos\theta_1,\cos\theta_2 | \mu_{\rm t},\sigma_{\rm t})$ is for spin orientation distribution, which is the mixture of isotropic and nearly aligned assemblies. $P_z(z|\gamma)$ is for the redshift distribution, where we assume the merger rate density evolving with redshift as $R(z)=R_0(1+z)^\gamma$, and $R_0$ is the local merger rate.

\subsection{model under stellar evolution theories (Alternative Model)}\label{app:models}

Since we find no evidence for a low-spin IMBH population (as presented in Section~\ref{sec:result}), the upper edge of the PIMG cannot be directly determined. We therefore adopt an alternative model in which the far-side edge is inferred under the guidance of stellar evolution theory, assuming that the low-spin subpopulation's cutoff defines the lower edge of the gap. In this model, we adopt a gap width of $80\pm8\,M_\odot$ based on simulations \citep{2020ApJ...902L..36F}, which found the width to remain approximately constant at $83^{+5}_{-8}\,M_\odot$, while the individual lower and upper edges are only weakly constrained. The theoretically motivated mass function is then written as
\begin{equation}
P_{m,1}(m|\Lambda_1) \propto \mathcal{PS}(m | \Lambda_1)\,f_{\rm Gap}(m|g_{\rm low},g_{\rm wide}),
\end{equation}
with 
\begin{equation}
f(m|g_{\rm low},g_{\rm wide}) = 
\begin{cases}
    1,  & \text{for } m < g_{\rm low}, \\
    0,       & \text{for } g_{\rm low} < m < g_{\rm low}+g_{\rm wide}, \\
    1,     & \text{for } m > g_{\rm low} +g_{\rm wide}
\end{cases}
\end{equation}

\subsection{mass-only model}\label{model:mass}

Even if the mass range of the PIMG in the first-generation black hole mass function is partly filled by higher-generation black holes from hierarchical mergers, a corresponding dip or deficit may still be imprinted on the overall binary black hole mass distribution. 
To search for this potential PIMG signature using mass information alone, we construct a population model that does not incorporate spin parameters. 
The parameterised joint mass distribution is given by
\begin{equation}
\pi(m_1,m_2|\Lambda) \propto P(m_1|\Lambda)\,P(m_2|\Lambda)\,(m_2/m_1)^\beta\,F_{\rm gap}(m_1,m_2|\Lambda),
\end{equation}
where $P(m|\Lambda)$ is the underlying (pre-pairing) mass function adopted for both component masses; here we use the \textsc{PowerLaw + 2 Peak} parametrisation introduced in \citet{2025arXiv250818083T}. 
The factor $F_{\rm gap}(m_1,m_2|\Lambda)$ describes the location of a possible gap or dip in the mass function of the secondary component (or of both components).

We consider three parametrisations of the gap/dip factor $F_{\rm gap}$.
The function $f_{\rm gap}$ describing a dip or gap in the mass distribution of a single component is defined as
\begin{equation}
f_{\rm gap}(m|g_{\rm low},g_{\rm up},A) = 
\begin{cases}
    1,  & \text{for } m < g_{\rm low}, \\[2pt]
    1-A, & \text{for } g_{\rm low} < m < g_{\rm up}, \\[2pt]
    1,  & \text{for } m > g_{\rm up},
\end{cases}
\end{equation}
where $g_{\rm low}$ and $g_{\rm up}$ mark the edges of the underlying gap or dip, and $A\in[0,1]$ controls the depth ($A=1$ corresponds to a completely empty gap; see \citet{2020ApJ...899L...8F}).
The three cases are then specified as follows.
\begin{itemize}
    \item \textbf{Case 1: Dip in both primary and secondary masses.}
    \begin{equation}
    \begin{aligned}
    &F_{\rm gap}(m_1,m_2|\Lambda) = \\
    &f_{\rm gap}(m_1|g_{\rm low},g_{\rm up},A)\,f_{\rm gap}(m_2|g_{\rm low},g_{\rm up},A).
    \end{aligned}
    \end{equation}
    \item \textbf{Case 2: Dip only in the secondary mass.}
    \begin{equation}
    \begin{aligned}
    &F_{\rm gap}(m_1,m_2|\Lambda) = \\
    &f_{\rm gap}(m_1|g_{\rm low},g_{\rm up},A{=}0)\,f_{\rm gap}(m_2|g_{\rm low},g_{\rm up},A).
    \end{aligned}
    \end{equation}
    \item \textbf{Case 3: Gap in the secondary mass function.}
    \begin{equation}
    \begin{aligned}
    &F_{\rm gap}(m_1,m_2|\Lambda) = \\
    &f_{\rm gap}(m_1|g_{\rm low},g_{\rm up},A{=}0)\,f_{\rm gap}(m_2|g_{\rm low},g_{\rm up},A{=}1).
    \end{aligned}
    \end{equation}
\end{itemize}
For all cases, the merger rate density evolution model is the same as in the main model introduced above.

\section{Results}\label{sec:result}
In this Section we display the results inferred with our main model, alternative model, as well as the mass-only model introduced in Section~\ref{sec:method}.

\begin{figure*}
	\centering  
\includegraphics[width=0.96\linewidth]{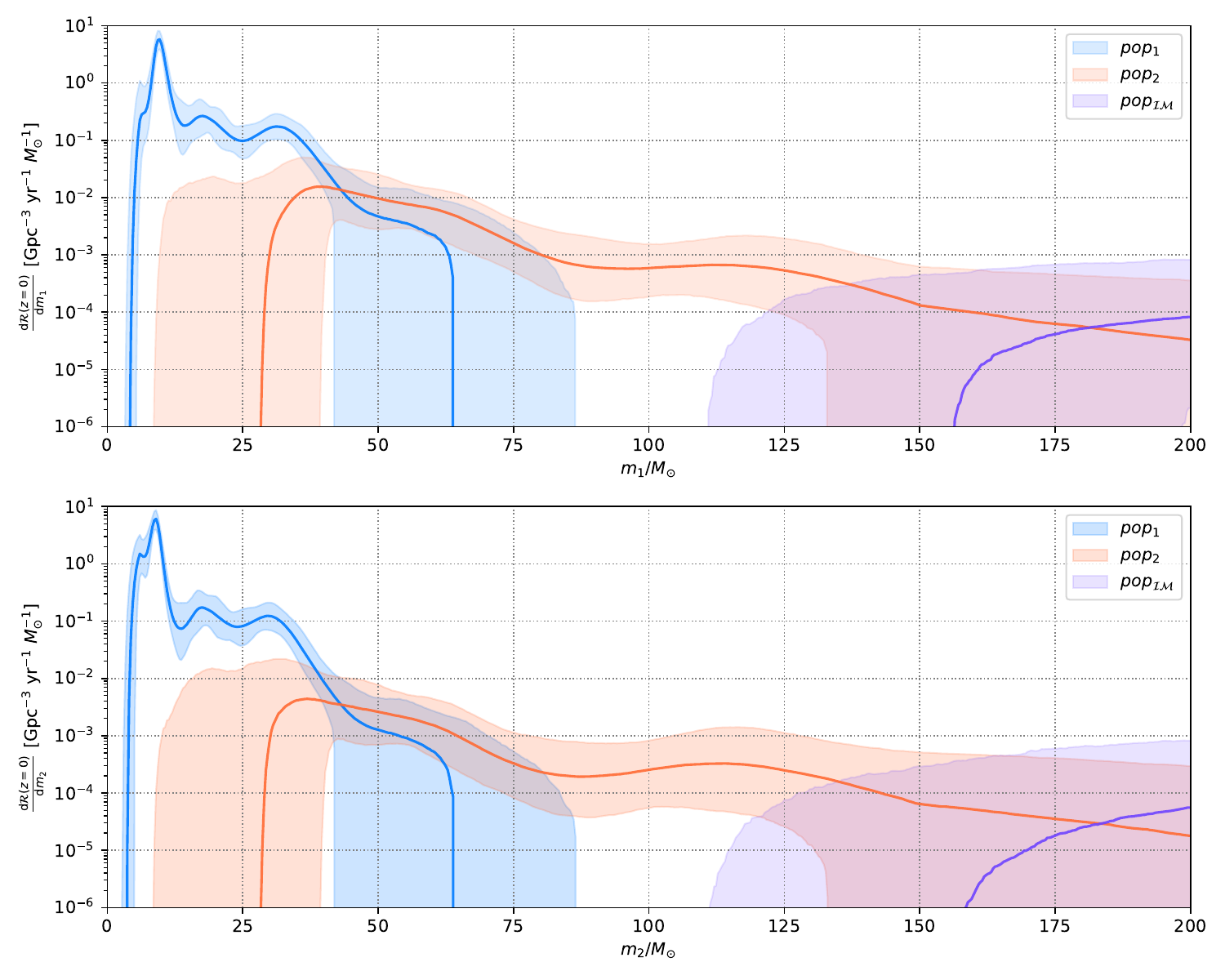}
\caption{Component mass distributions of the primary (top) and secondary (bottom) black holes for the three subpopulations inferred with the Main Model. The solid curves are the medians and the shaded regions are for the 90\% credible intervals.}
\label{fig1}
\end{figure*}

\begin{figure}
	\centering  
\includegraphics[width=0.98\linewidth]{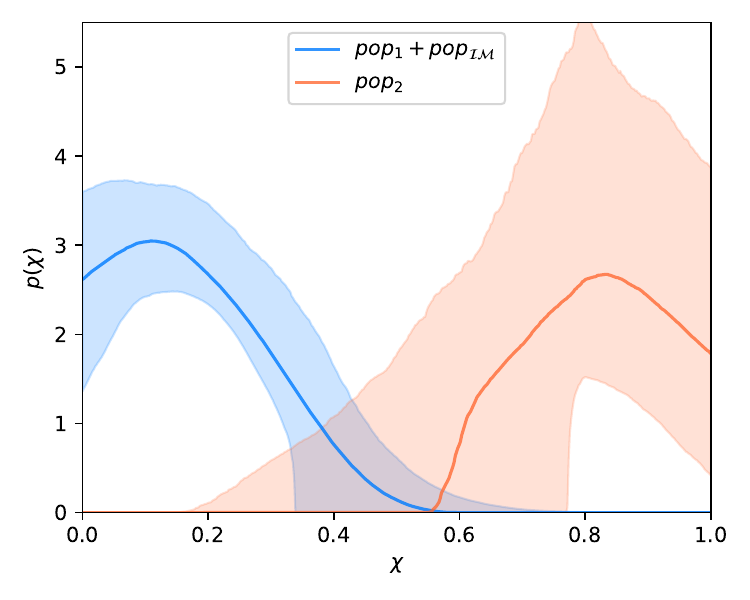}
\caption{Distribution of spin magnitudes of black holes for the subpopulations inferred with the Main Model. The solid curves are the medians and the shaded regions are for the 90\% credible intervals.}
\label{fig2}
\end{figure}

\begin{figure*}
	\centering  
\includegraphics[width=0.96\linewidth]{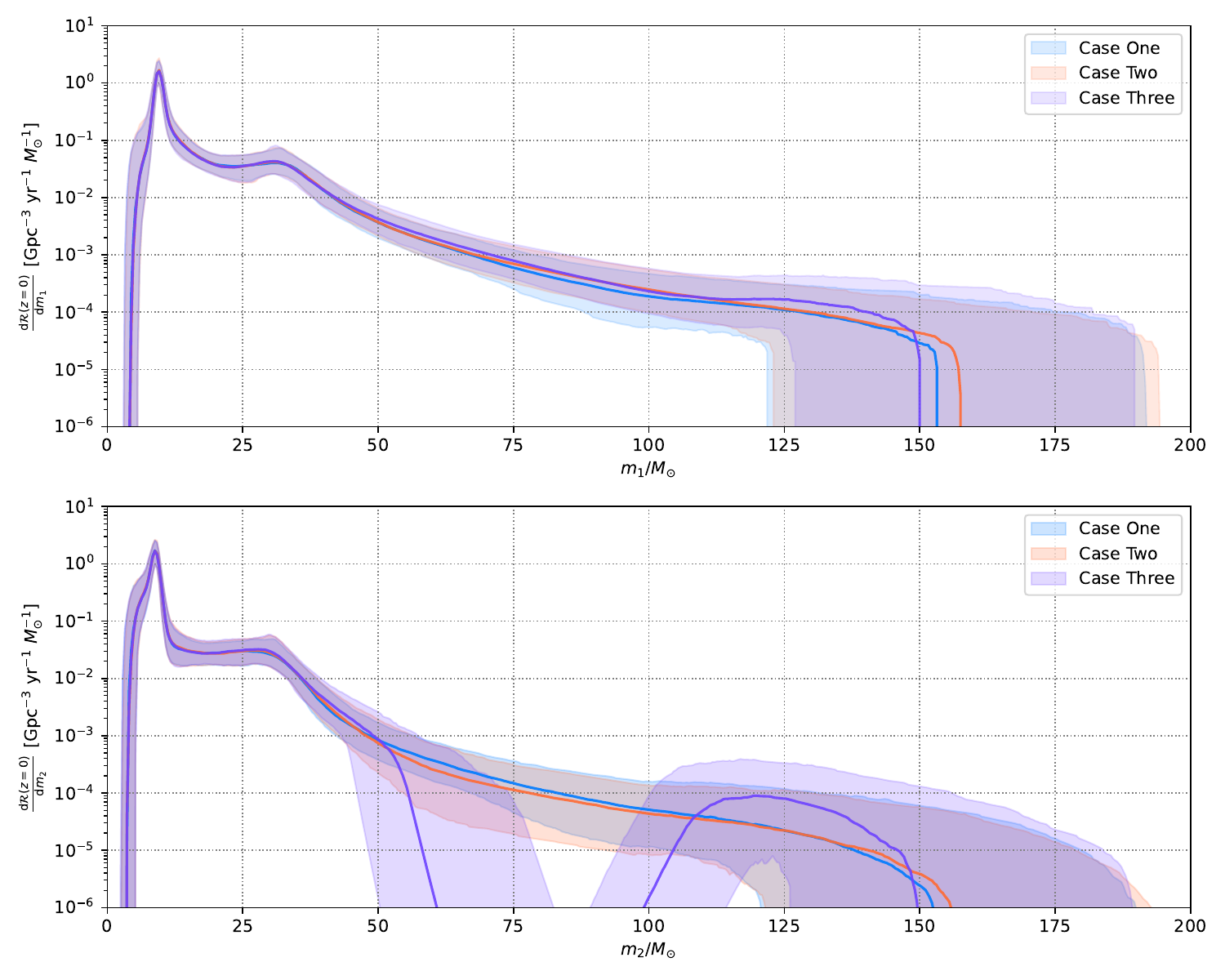}
\caption{Mass distributions of the primary (a) and secondary (b) black holes inferred with the mass-only model for three cases. The solid curves are the medians and the shaded regions are for the 90\% credible intervals.}
\label{fig3}
\end{figure*}

\begin{figure*}
	\centering  
\includegraphics[width=0.48\linewidth]{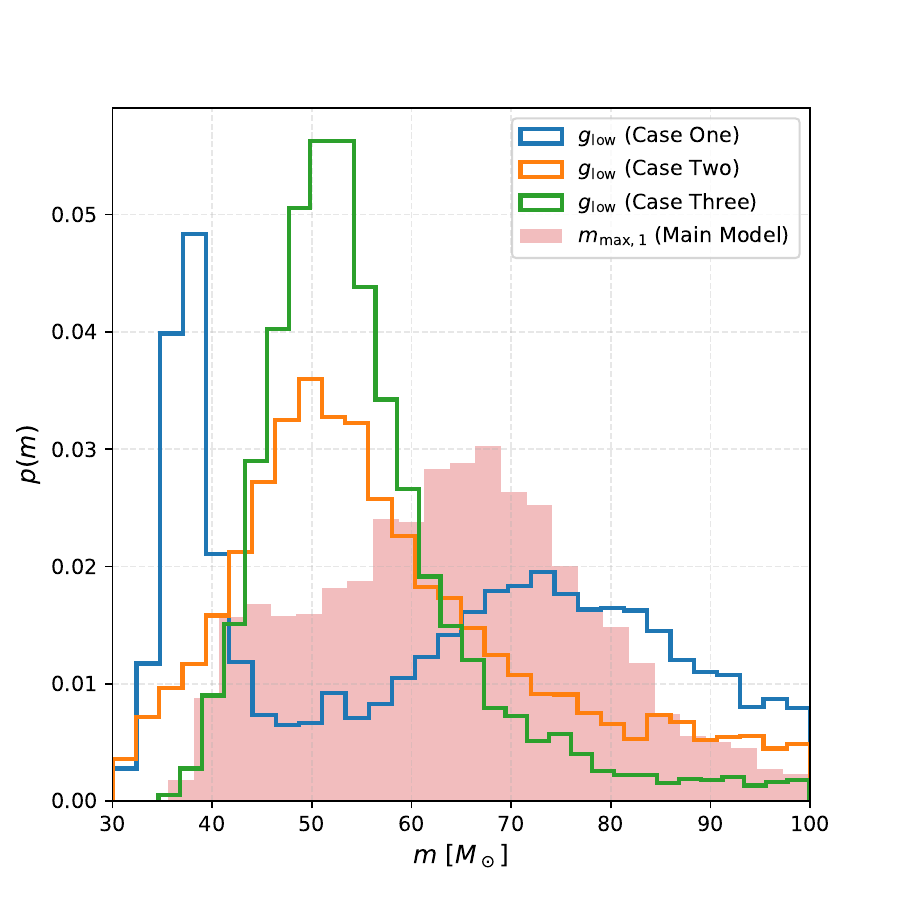}
\includegraphics[width=0.48\linewidth]{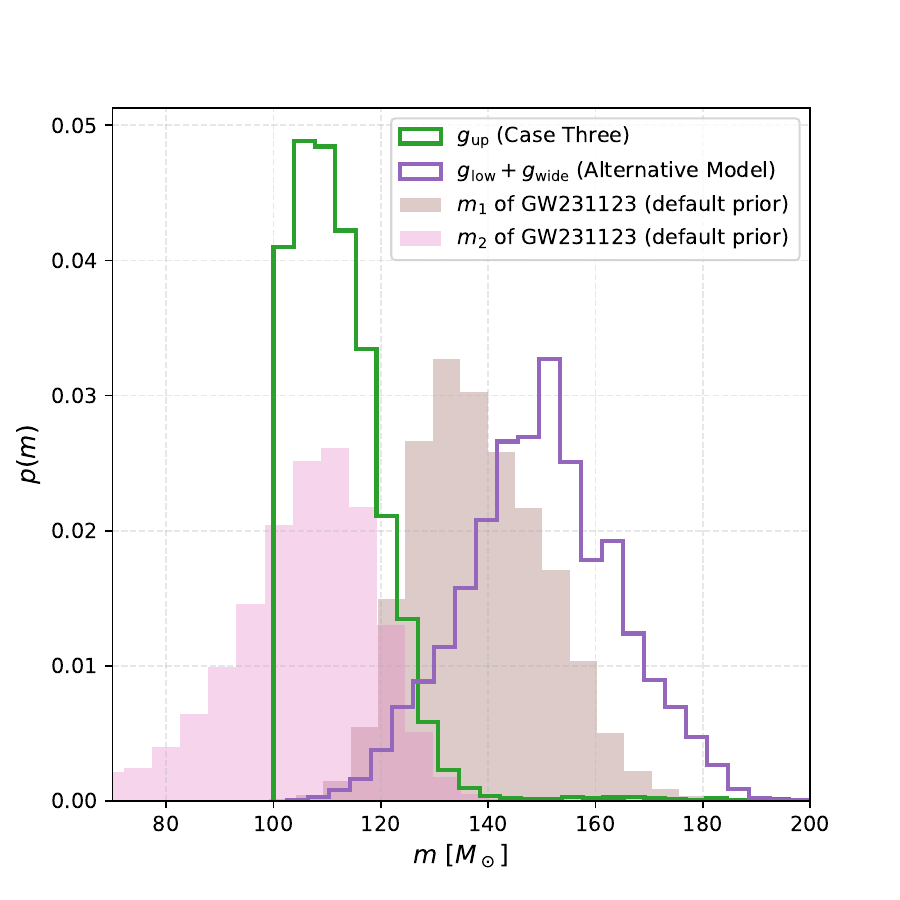}
\caption{Left: The posterior distributions of maximum masses that are potentially associated to the lower edge of PIMG. Right:The posterior distributions of cutoff masses that are potentially associated to the upper edge of PIMG, compared to the component masses of GW231123.}
\label{fig:mmax}
\end{figure*}

\subsection{No evidence for low-spin IMBHs}

Figures~\ref{fig1} and \ref{fig2} show the mass and spin distributions of the BBH population inferred with our main model. 
Two distinct subpopulations are clearly identified in both the primary and secondary black holes, indicated by the blue and orange curves.

The first subpopulation, referred to as the low-spin group, is characterized by small dimensionless spin magnitudes ($\chi \lesssim 0.3$), peaking at $\chi \sim 0.15$. 
Its mass distribution is confined to the lower-mass range ($\sim\!5$--$65\,M_\odot$) and exhibits a pronounced cutoff above $65^{+23}_{-22}\,M_\odot$. 
This behaviour is consistent with the expectations for black holes formed through stellar core collapse and the upper mass limit imposed by (pulsational) pair-instability processes \citep{2019ApJ...887...53F,2020ApJ...902L..36F,2020ApJ...888...76M}.

In contrast, the second (high-spin) subpopulation shows significantly larger spin magnitudes ($\chi \gtrsim 0.6$), with a peak around $\chi \sim 0.8$. 
Its mass distribution is broad, extending from about $20\,M_\odot$ up to $\sim\!150\,M_\odot$ and possibly beyond, in agreement with expectations for hierarchical merger products.

Crucially, we find no evidence for an additional low-spin, high-mass subpopulation above $100\,M_\odot$, 
as the mixture fraction $r_{\mathcal{IM}}$ of the $\mathcal{PL}$ component in Eq.~(\ref{equation3})—introduced to capture potential first-generation IMBHs—is consistent with zero. 
If we interpret the low-spin subpopulation as first-generation BHs, the 90\% credibility upper limit on the local merger rate of collapse-formed IMBH mergers is $< 0.077\,\mathrm{Gpc}^{-3}\,\mathrm{yr}^{-1}$, see Appendix~\ref{app:results}.

\subsection{The far-side edge of mass gap}

We find no evidence for a low-spin subpopulation with masses above $100\,M_\odot$, which precludes a direct measurement of the PIMG's upper edge. 
To address this, we employ the Alternative Model described in Section~\ref{model:mass}. 
Adopting a Gaussian prior on the PIMG width of $80\pm8\,M_\odot$ ($1\sigma$) \citep{2020ApJ...902L..36F}, we indirectly constrain the upper edge to be $g_{\rm up} = 150\pm24\,M_\odot$ (see Figure~\ref{fig:Gap_post} in the Appendix).

We therefore analyze the PIMG using the mass-only model (Section~\ref{model:mass}). 
This approach is motivated by the expectation that if core-collapse IMBHs are significantly more abundant than those formed via hierarchical mergers, a distinct signature of the PIMG's upper edge should be imprinted on the mass function.
Figure~\ref{fig3} shows the inferred primary and secondary mass distributions for the three cases. 
In the primary mass distribution, we find no evidence for a pronounced dip or gap, with all models yielding nearly identical results. 
This consistency suggests that hierarchical mergers are sufficiently frequent in the high-mass regime to largely erase the expected PIMG signature. 
Regarding the secondary mass distribution, we similarly find no statistically significant evidence for a dip or gap. 
While an entirely empty gap ($A=1$) is mildly preferred over a completely filled one ($A=0$), the depth of any potential dip remains weakly constrained (see Figure~\ref{app:corner} in Appendix~\ref{app:results}).

For Case 3 (empty gap in the secondary-mass distribution), we obtain $g_{\rm low} = 53^{+23}_{-10}\,M_\odot$ and $g_{\rm up} = 111^{+15}_{-9}\,M_\odot$. 
The lower edge $g_{\rm low}$ is consistent with results from other parametric models~\citep{2025arXiv250904151T, 2025arXiv251018867R} and agrees with $m_{\rm max,1}$ (i.e., $g_{\rm low}$) in the Main (Alternative) Model. 
However, the upper edge $g_{\rm up}$ is notably lower than the value inferred from $g_{\rm low}+g_{\rm wide}$ in the Alternative Model (see Figure~\ref{fig:mmax}) and than theoretical expectations~\citep{2020ApJ...902L..36F}, revealing some tension. 
We note that $g_{\rm up}$ is primarily driven by the secondary mass of GW231123, which likely accounts for this discrepancy.

\subsection{The origin of LVK's most massive black holes and their merger rate}

Analysis of the Main Model indicates that both components of GW231123 belong to the high-spin subpopulation, consistent with a hierarchical merger scenario \citep{2017PhRvD..95l4046G, 2017ApJ...840L..24F, 2025arXiv250708219T}. Moreover, their masses do not lie significantly beyond the theoretically expected upper edge of the PIMG obtained from the Alternative Model \citep{2020ApJ...902L..36F} (Figure~\ref{fig:mmax}). Thus, GW231123 is most likely a binary composed of higher-generation black holes.

We further find that the maximum masses of both subpopulations remain essentially unconstrained. Theoretically, each could extend to $\gtrsim 300\,M_\odot$, and the absence of such detections is likely due to the limited sensitivity of current detectors \citep{2015CQGra..32g4001L,2018LRR....21....3A}. Accordingly, we place a model-dependent 90\% credible upper limit on the merger rate density of IMBHs with $m_1 \gtrsim 150\,M_\odot$:
$
{{\rm d}\mathcal{R}(z=0)}/{{\rm d}m_1} \lesssim 10^{-4} \text{--} 10^{-3}~\mathrm{Gpc}^{-3}\,\mathrm{yr}^{-1}\,M_\odot^{-1}
$
(see Figure~\ref{fig:upperlimit}). Detecting such systems will likely require next-generation observatories \citep{2023LRR....26....2A,2025RPPh...88e6901L,2025arXiv250312263A}.

\section{Conclusions and Discussion}
\label{sec:diss}

We conducted population analysis of binary black holes in GWTC-4, focusing on their component masses and spins to investigate the origin of massive black holes (within or beyond the expected PIMG). Our main findings are:
\begin{enumerate}
    \item \textbf{No evidence for low-spin (stellar-collapse) IMBHs:} We find no significant evidence for a subpopulation of low-spin IMBHs that would be indicative of formation via direct stellar core collapse. The mixture fraction for such a potential population is constrained to $<0.017$ (90\% credibility). This places an upper limit on the local merger rate of collapse-formed IMBHs at $0.077~\mathrm{Gpc}^{-3}~\mathrm{yr}^{-1}$.
    \item \textbf{Constraints on the PISN mass gap (PIMG):} The mass distribution of the identified low-spin subpopulation, consistent with first-generation stellar-origin BHs, shows a truncation. We constrain the \textit{lower} edge of the pair-instability mass gap to $\sim 65 M_\odot$, though find no statistically significant evidence in the current data to determine the \textit{upper} boundary of the PIMG. There may be a gap with \textit{upper} edge (of $111^{+15}_{-9}M_\odot$) in the secondary-mass distribution, which, however, may not be the PIMG.
    If we assume the width of PIMG to be $80\pm8M_\odot$ as predicted by \citet{2020ApJ...902L..36F}, we find the \textit{upper} edge to be $150\pm24M_\odot$.
    
    \item \textbf{Hierarchical merger origin for the most massive BHs:} The IMBHs observed (e.g., in GW231123) belong to the high-spin ($\chi \gtrsim 0.6$) subpopulation. Their spin and mass distributions are consistent with expectations for BHs formed through successive hierarchical mergers in dense environments, like globular clusters or AGN disks. 
\end{enumerate}

While hierarchical mergers in dense environments remain the most consistent formation channel for the most massive BBHs in our findings, alternative origins for high-spin, high-mass black holes cannot be entirely excluded.  
For instance, {Primordial Black Holes (PBHs)} could acquire high spins through accretion \citep{2025arXiv250715701Y, 2025arXiv250809965D}, although their abundance and merger rates are still highly uncertain \citep{2017PhRvD..96l3523A, 2020JCAP...11..028D}.  
{Population III stars or chemically homogeneous evolution (CHE)} at low metallicity can naturally produce rapidly rotating progenitors and give rise to high-mass, high-spin BBHs such as GW231123 \citep{2025ApJ...994L..54P, 2025ApJ...995L..76P} \footnote{Lensing and Overlapping effects may also produce GW231123-like events \citep[e.g.]{2025arXiv251217550H,2025arXiv251220890Y,2026arXiv260414247H}}.  
{Stellar mergers or sustained accretion} onto stellar-mass BHs \citep{2020ApJ...904L..13R,2025ApJ...994L..37K,2025arXiv250808558B} could also form massive, highly spinning BHs (but see \citet{2022ApJ...938...45B}).  
{Direct collapse} of a single rotating massive star can produce BHs resembling the component objects of GW231123 \citep{2025ApJ...993L..54G,2025arXiv250810088C,2025PhRvL.135s1401B}.  
These alternative pathways make it difficult to distinguish hierarchical merger products from other formation scenarios using the currently available BBH catalog, which in turn prevents a clean, model-independent determination of the far-side edge of the PIMG from the existing population.  
Nevertheless, a clear observational diagnostic can break this degeneracy: the robust detection of even a single IMBH merger event with low spin would provide definitive evidence for IMBHs formed through stellar core collapse, thereby directly constraining the upper edge of the PIMG. 

\citet{2021ApJ...909L..23E} demonstrated that the upper edge of the PIMG could, in principle, be constrained to the percent level with a few tens of events. However, a substantial population of hierarchical mergers in the relevant mass range may smear out this sharp cutoff, obscuring the mass boundary expected from stellar evolution. Such an effect has already been observed for the \textit{lower} edge of the PIMG \citep{2023PhRvX..13a1048A} and was addressed in previous work \citep{2024PhRvL.133e1401L}. Future low-frequency observatories such as LISA, TianQin, and Taiji \citep{2016CQGra..33c5010L, 2017arXiv170200786A, 2017NSRev...4..685H} will therefore be crucial: by detecting the slowly spinning IMBH binaries expected from core collapse, they could directly reveal this boundary \citep{2016PhRvL.116w1102S, 2019MNRAS.488L..94M}.

Larger samples of detections, combined with more sophisticated population models that incorporate masses, spins, and redshifts, will be essential to disentangle the formation channels of the most massive binaries and to explore their potential role as seeds for supermassive black holes \citep{2005MNRAS.357..275K}. Furthermore, as distinct black hole subpopulations are identified, they will also contribute to cosmological studies through ``spectral sirens'' or ``multi-spectral sirens'' \citep{2019ApJ...883L..42F, 2021ApJ...908..215Y, 2024ApJ...976..153L}. For instance, the core-collapse IMBH population -- and thus the location of the PIMG's upper edge -- could provide novel constraints on cosmic expansion \citep{2022PhRvL.129f1102E}.


\vspace{5mm}

\textbf{\it Acknowledgments:}
We thank Yi-Zhong Fan, Shao-Peng Tang, and Yin-Jie Li for helpful suggestions. This work is supported by the National Natural Science Foundation of China (NSFC, No. 12303063 and No. 12203101). Y.Qin is supported by NSFC (No. 12473036 and No. 12573045) and Anhui Provincial Natural Science Foundation (No. 2308085MA29). This research has made use of data and software obtained from the Gravitational Wave Open Science Center (https://www.gw-openscience.org), a service of LIGO Laboratory, the LIGO Scientific Collaboration and the Virgo Collaboration. LIGO is funded by the U.S. National Science Foundation. Virgo is funded by the French Centre National de Recherche Scientifique (CNRS), the Italian Istituto Nazionale della Fisica Nucleare (INFN) and the Dutch Nikhef, with contributions by Polish and Hungarian institutes.

\vspace{5mm}

\software{Bilby \citep[version 1.1.4, ascl:1901.011, \url{https://git.ligo.org/lscsoft/bilby/}]{2019ascl.soft01011A},
          PyMultiNest \citep[version 2.11, ascl:1606.005, \url{https://github.com/JohannesBuchner/PyMultiNest}]{2016ascl.soft06005B}.
          }


\appendix
\section{Hierarchical Bayesian Inference}\label{app:meth}
Following the framework of \citet{2019MNRAS.486.1086M,2021ApJ...913L...7A,2023PhRvX..13a1048A}, given the population hyperparameters $\Lambda$, the likelihood of the GW data $\{d\}$ from $N_{\rm det}$ detections is
\begin{equation}
\mathcal{L}(\{d\}|\Lambda)\propto N^{N_{\rm det}}\,e^{-N_{\rm exp}} \prod_{i=1}^{N_{\rm det}} \int \pi(\theta_i|\Lambda)\,\mathcal{L}(d_i|\theta_i)\,d\theta_i,
\end{equation}
where $N=\int R(z|\Lambda)\,\frac{dV_c}{dz}\,\frac{T_{\rm obs}}{1+z}\,dz$ is the total number of mergers occurring in the surveyed spacetime volume, and $N_{\rm exp}=N\int P({\rm det}|\theta)\,\pi(\theta|\Lambda)\,d\theta$ is the expected number of detections, with $P({\rm det}|\theta)$ the detection probability. 
$N_{\rm exp}$ can be computed via a Monte Carlo integral over a reference set of injected signals\footnote{Obtained from \url{https://zenodo.org/doi/10.5281/zenodo.5636815}.}, and the individual-event likelihood $\mathcal{L}(d_i|\theta_i)$ is evaluated using the available posterior samples; see the Appendix of \citet{2021ApJ...913L...7A} for details.
Following \cite{2025arXiv250818083T}, the uncertainty of likelihood raising from the Monte Carlo integrals  \citep{2023PhRvX..13a1048A,2023MNRAS.526.3495T} are controlled to be $\sigma^2_{\mathcal{L}}<1$.

\section{Additional results}\label{app:results}

\begin{figure}
	\centering  
\includegraphics[width=0.48\linewidth]{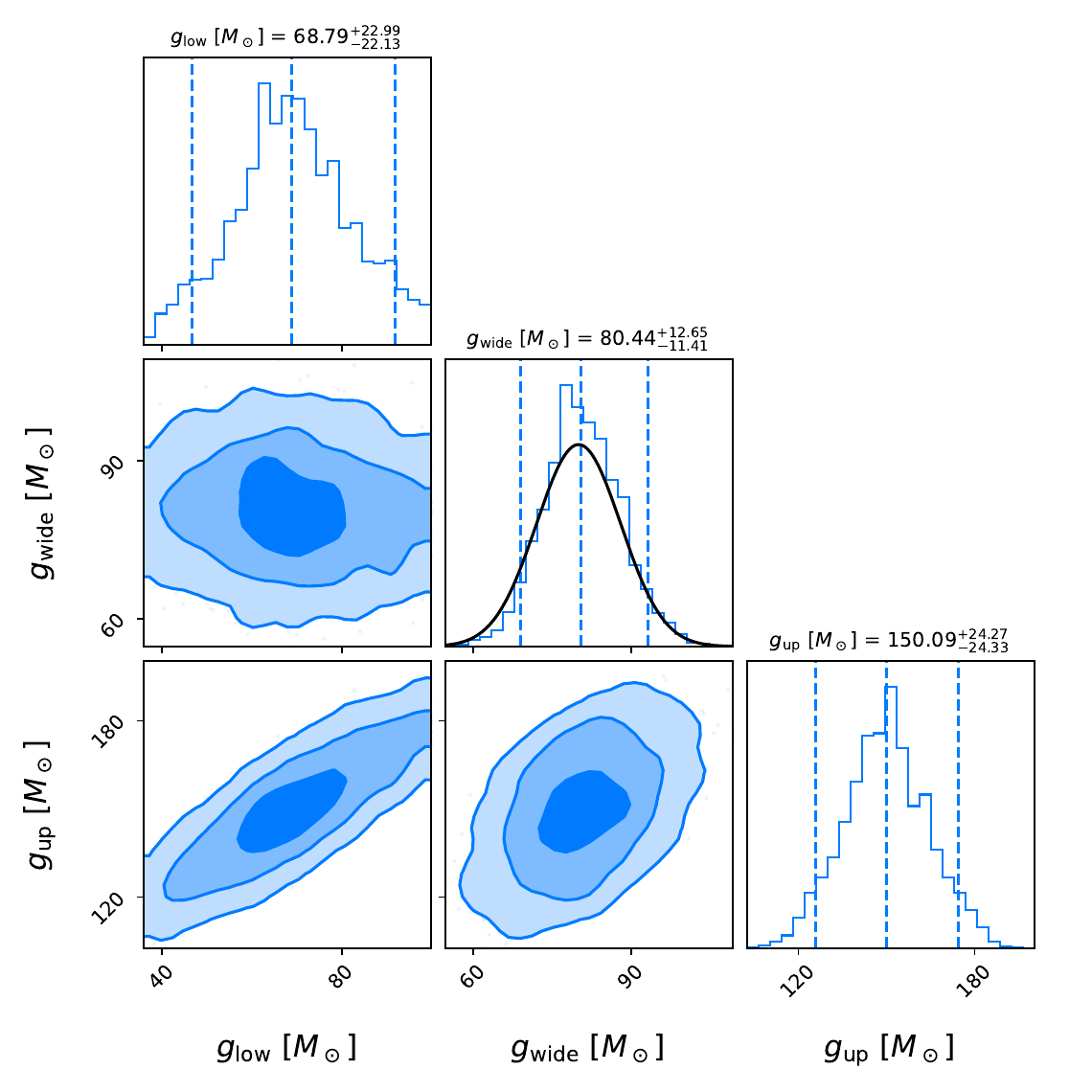}
\caption{The posterior distributions of $g_{\rm low}$, $g_{\rm wide}$, and $g_{\rm up} \equiv g_{\rm low} + g_{\rm wide}$ — which describe the edges and width of the PIMG — as obtained with the Alternative Model. The black curve indicates the prior on the PIMG width, taken to be $80\pm8\,M_\odot$ following \cite{2020ApJ...902L..36F}. The vertical lines indicate the median values and 90\% credible intervals.}
\label{fig:Gap_post}
\end{figure}

Figure~\ref{fig:Gap_post} shows the constraints on the PIMG features from the Alternative Model, namely $g_{\rm low}$, $g_{\rm wide}$, and the derived upper edge $g_{\rm up} \equiv g_{\rm low} + g_{\rm wide}$. In this model we adopt a theoretically motivated prior on the PIMG width of $g_{\rm wide} = 80 \pm 8\,M_\odot$, based on the simulations of \citet{2020ApJ...902L..36F}, which indicate that the PIMG width remains approximately constant at $83^{+5}_{-8}\,M_\odot$, while the lower edge is poorly constrained, varying between about $40$ and $90\,M_\odot$.
With this Alternative Model, we obtain $g_{\rm low} = 68.8^{+23.0}_{-22.1}\,M_\odot$, consistent with the result from the Main Model, and the upper edge of the PIMG is constrained to be $g_{\rm up} = 150.1 \pm 24.3\,M_\odot$.

\begin{figure}
	\centering  
\includegraphics[width=0.48\linewidth]{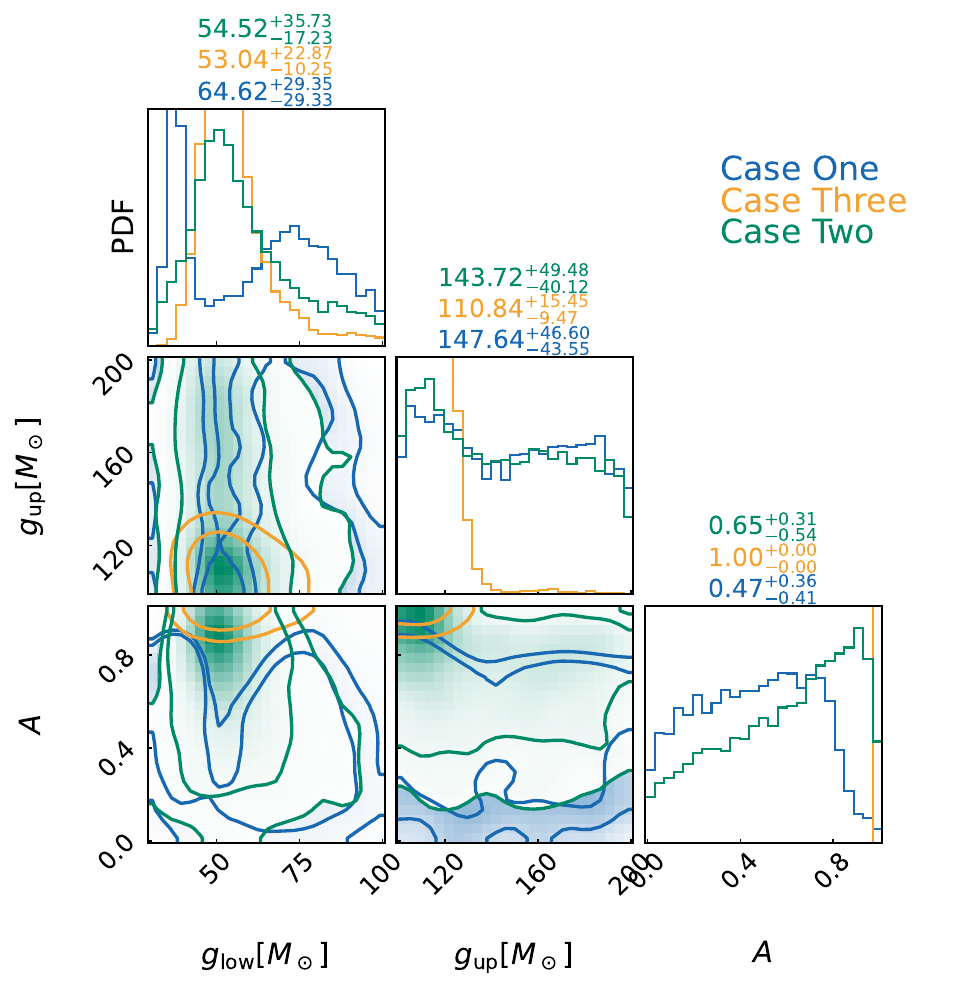}
\caption{The posterior distribution of hyperparameters obtained with the mass-only models (Case One, Case Two, and Case Three). The values are for median and 90\% credible intervals.}
\label{app:corner}
\end{figure}

Table~\ref{tab:BF} gives the Bayes factors of the three mass-only model cases with respect to Case One, showing that all three are competitive.
Figure~\ref{app:corner} shows the posterior distributions; the inferred gap edges are broadly consistent across cases. 
For Case~3 (empty gap in the secondary-mass distribution), we obtain $g_{\rm low} = 53^{+23}_{-10}\,M_\odot$ and $g_{\rm up} = 111^{+15}_{-9}\,M_\odot$. 
The lower edge agrees with other parametric models~\citep{2025arXiv250904151T, 2025arXiv251018867R}, with $m_{\rm max,1}$ in the Main/Alternative Models, and with the drop in the primary-mass function observed in a previous catalog~\citep{2021ApJ...913...42W}.
The upper edge, however, is significantly lower than both the value from $g_{\rm low}+g_{\rm wide}$ in the Alternative Model (Figure~\ref{fig:mmax}) and theoretical expectations~\citep{2020ApJ...902L..36F}, indicating tension. 

\begin{figure}
	\centering  
\includegraphics[width=0.48\linewidth]{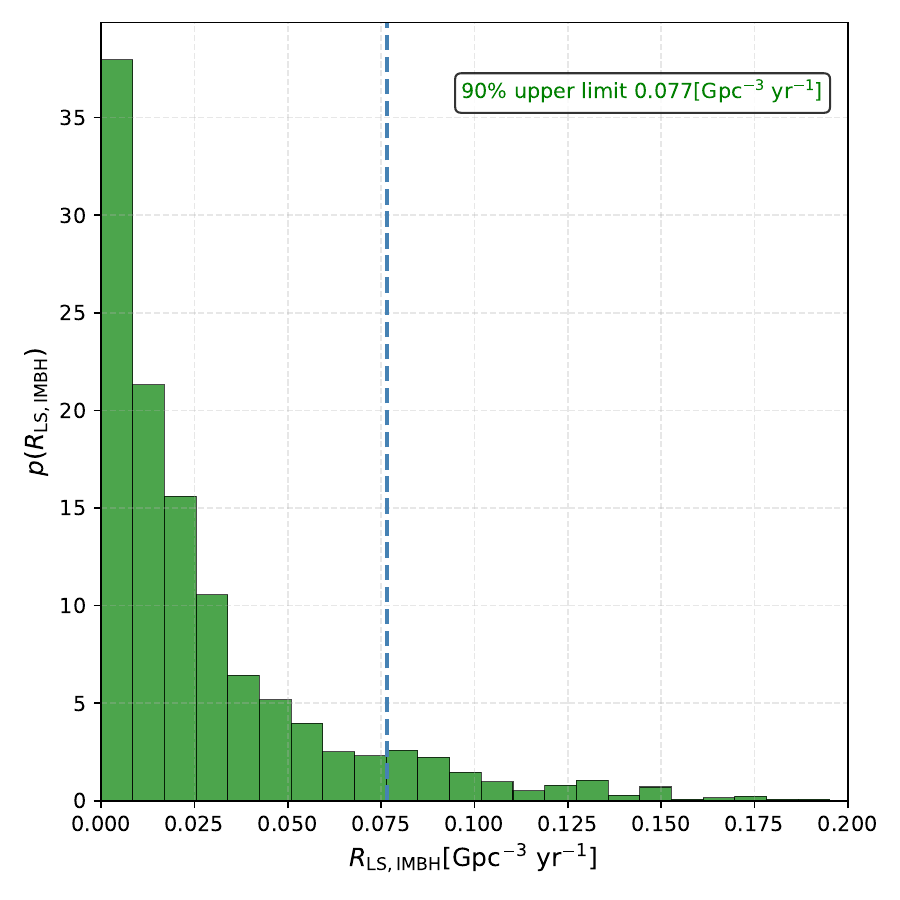}
\caption{Merger rate of the low-spin IMBHs inferred with the Main Model. The vertical line indicate the 90\% upper limit.}
\label{fig:Rate}
\end{figure}

\begin{figure}
	\centering  
\includegraphics[width=0.48\linewidth]{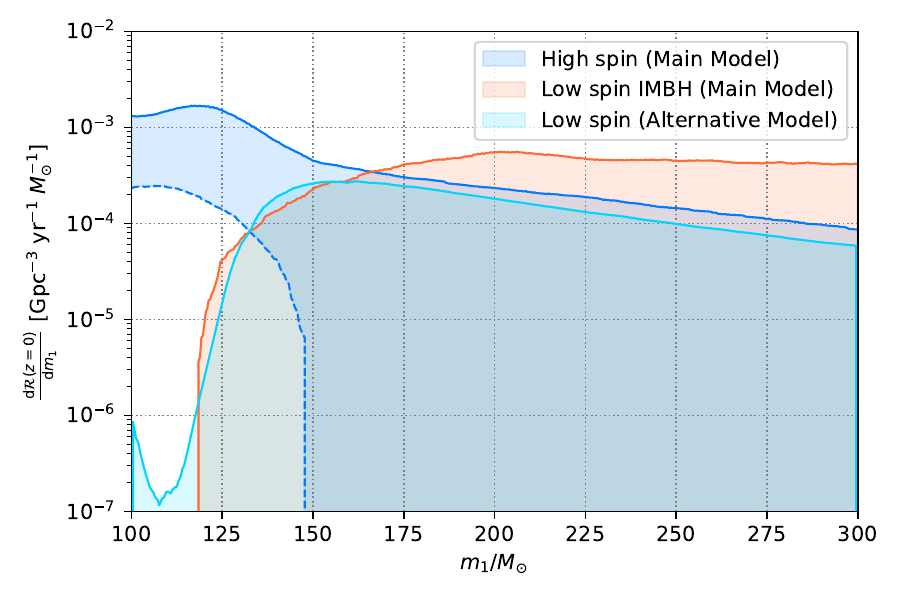}
\caption{Upper limit (90\%) of the IMBH mergers from different formation channels.}
\label{fig:upperlimit}
\end{figure}

Figure~\ref{fig:Rate} shows the merger rate density inferred for the low-spin IMBH subpopulation under the Main Model, which is consistent with zero across the analyzed mass range, with a 90\% upper limit of $0.077~\mathrm{Gpc}^{-3}\,\mathrm{yr}^{-1}$.
Figure~\ref{fig:upperlimit} displays the corresponding upper limits on the merger rate density for IMBHs from different subpopulations; we find the 90\% credible upper limit on the merger rate density of IMBHs with $m_1 \gtrsim 150\,M_\odot$ to be
$\mathrm{d}\mathcal{R}(z=0)/\mathrm{d}m_1 \lesssim 10^{-4} \text{--} 10^{-3}~\mathrm{Gpc}^{-3}\,\mathrm{yr}^{-1}\,M_\odot^{-1}$.
Detecting such systems will likely require next-generation observatories \citep{2023LRR....26....2A,2025RPPh...88e6901L,2025arXiv250312263A}.

\begin{table*}[htpb]
\centering
\caption{Summary of model parameters.}\label{tab:prior}
\begin{tabular}{lcccc}
\hline
\hline
Parameter     &  Description & Prior \\
\hline
\hline
\multicolumn{3}{c}{\bf \textsc{PowerLaw+Spline} mass function}\\
$m_{\rm min,i}[M_{\odot}]$   & The minimum mass & $U(2,50)$  \\
$m_{\rm max,1}[M_{\odot}]$ / $m_{\rm max,2}[M_{\odot}]$   & The maximum mass & $U(20,100)$ / $U(20,500)$  \\
$\alpha_i$ & Slope index of the power-law mass function & $U(-8,8)$ \\
$\delta_{\rm m,i}[M_{\odot}]$ & Smooth scale of the mass lower edge & $U(0,10)$ \\
$\{f_i^i\}_{j=2}^{11}$ & Interpolation values of perturbation function & $\mathcal{N}(0,1)$ \\
$r_2$ & mixture fraction for the second subpopulation & $U(0,1)$ \\
constraints & & $m_{\rm min,i}<m_{\rm max,i}$ \\
$\beta_{q}$ & Slope index of the mass-ratio distribution & $U(-8,8)$ \\
\hline
\multicolumn{3}{c}{\bf Spin distribution}\\
$\chi_{{\rm min},1}$ / $\chi_{{\rm min},2}$  & Lower edge for $\chi$ distribution  & 0 / $U(0,0.8)$  \\
$\chi_{{\rm max},1}$ / $\chi_{{\rm max},2}$  & Upper edge for $\chi$ distribution & $U(0.2,1)$ / 1  \\
$\mu_{\chi ,i}$ & Center values for $\chi$ distribution  & $U(0,1)$ \\
$\sigma_{\chi ,i}$ & Width of the $\chi$ distribution  & $U(0.05, 0.5)$ \\
constraints & & $\mu_{\chi,1}<\mu_{\chi,2}$ \\
$\mu_{t}$ & peak of the $\cos\theta_{1,2}$ distribution for Gaussian& $U(0.1, 4)$ \\
$\sigma_{t}$ & width of the $\cos\theta_{1,2}$ distribution for Gaussian& $U(0.1, 4)$ \\
$\zeta$ & the mixture fraction for Gaussian & $U(0,1)$ \\
\hline
\multicolumn{3}{c}{\bf Rate evolution model} \\
$\lg (R_0[{\rm Gpc}^{-3}~{\rm yr}^{-1}])$ & Local merger rate  density & $ U(-3,3)$ \\
$\gamma$ & Slope of the power-law & $U(-8,8)$ \\
\hline
\multicolumn{3}{c}{\bf Special for the Main $\textsc{Model}$} in the text\\
$m_{{\rm min},\mathcal{IM}}[M_{\odot}]$  & The minimum mass of stellar-formed IMBHs ($pop_\mathcal{IM}$) & $U(100,150)$  \\
$m_{{\rm max},\mathcal{IM}}[M_{\odot}]$   & The maximum mass for $pop_\mathcal{IM}$ & $U(100,500)$  \\
$\alpha_\mathcal{IM}$ & Slope index for $pop_\mathcal{IM}$ & $U(-8,8)$ \\
$r_\mathcal{IM}$ & Mixture fraction for $pop_\mathcal{IM}$ in the first subpopulation & $U(0,1)$ \\
constraints & & $m_{{\rm min},\mathcal{IM}}<m_{{\rm max},\mathcal{IM}}$ \\
\hline
\multicolumn{3}{c}{\bf Special for the Alternative Model }\\
$g_{\rm low}[M_{\odot}]$  & The lower edge of PIMG & $U(30,100)$  \\
$g_{\rm wide} [M_{\odot}]$   & The width of PIMG & $\mathcal{N}(80,8)$  \\
\hline
\hline
\end{tabular}
\\
\begin{tabular}{l}
Note: $U$, $LogU$, $\mathcal{N}$ are for Uniform, Log-Uniform, Normal distribution.
\end{tabular}
\end{table*}

\begin{table*}[htpb]
\centering
\caption{Summary of parameters for mass-only model.}\label{tab:mass_prior}
\begin{tabular}{lcccc}
\hline
\hline
Parameter     &  Description & Prior \\
\hline
$m_{\rm min}[M_{\odot}]$   & The minimum mass & $U(2,10)$  \\
$m_{\rm max}[M_{\odot}]$   & The maximum mass & $U(100,200)$ \\
$\alpha$ & Slope index of the power-law mass function & $U(-8,8)$ \\
$\delta_{\rm m}[M_{\odot}]$ & Smooth scale of the mass lower edge & $U(0,10)$ \\
$\mu_{1}[M_{\odot}]$ & Center value for the first peak  & $U(5,15)$ \\
$\sigma_{1}[M_{\odot}]$ & Width of the first peak  & $U(1, 10)$ \\
$\mu_{2}[M_{\odot}]$ & Center value for the second peak  & $U(15,50)$ \\
$\sigma_{2}[M_{\odot}]$ & Width of the second peak  & $U(1, 10)$ \\
$r_p$ & mixture fraction for the Gaussian peaks & $U(0,1)$ \\
$r_2$ & mixture fraction of the second Gaussian peak & $U(0,1)$ \\
$\beta_{q}$ & Slope index of the pairing function & $U(-8,8)$ \\
$g_{\rm low}[M_{\odot}]$  & The lower edge of PIMG & $U(30,100)$  \\
$g_{\rm up} [M_{\odot}]$   & The upper edge of PIMG & $\mathcal{N}(100,200)$  \\
\hline
\hline
\end{tabular}
\\
\begin{tabular}{l}
Note: $U$, $LogU$, $\mathcal{N}$, $\mathcal{G}$ are for Uniform, Log-Uniform, Normal distribution, and Gaussian distribution.
\end{tabular}
\end{table*}

\begin{table}[htpb]
\centering
\caption{Bayes factors for mass-only models relative to Case One}\label{tab:BF}
\begin{tabular}{c|c}
\hline\hline
 Cases for mass-only model & $\ln\mathcal{B}$ \\
\hline
Case One & 0 \\
Case Two & 0.4 \\
Case Three & 0.3 \\
\hline\hline
\end{tabular}
\end{table}

\clearpage
\bibliography{export-bibtex}{}
\bibliographystyle{aasjournal}

\end{CJK*}
\end{document}